# Transient Nature of Negative Capacitance in Ferroelectric Field-Effect Transistors

Kwok Ng[1], Steven J. Hillenius[1], and Alexei Gruverman[2]

[1] Semiconductor Research Corporation, Durham, NC 27703, USA
[2] Department of Physics and Astronomy, University of Nebraska-Lincoln, NE 68588, USA

*Abstract*—Negative capacitance (NC) in ferroelectrics, which stems from the imperfect screening of polarization, is considered a viable approach to lower voltage operation in the field-effect transistors (FETs) used in logic switches. In this paper, we discuss the implications of the transient nature of negative capacitance for its practical application. It is suggested that the NC effect needs to be characterized at the proper time scale to identify the type of circuits where functional NC-FETs can be used effectively.

Recently, there has been much interest in the negative capacitance (NC) phenomenon in ferroelectrics, which could potentially reduce the energy consumption in field-effect transistors (FETs) due to the enhanced subthreshold slope [1-3]. While there is much debate on the physics, interpretation, and applications of this NC effect, there is a general consensus that the phenomenon is metastable in nature [4]. In order to measure negative capacitance, as well as to put this phenomenon to use in a real device, particular constraints have to be considered. In particular, Catalan et al [5] pointed out the transient effects associated with the dynamics of screening processes in ferroelectric capacitors. We follow up on that discussion and address the transient nature of NC-FET functionality that must follow.

One of the main problems in transistor scaling is the power consumption, which turns into heat of the integrated circuits. In MOSFETs, it is critical to maximize the subthreshold slope so that a certain ON/OFF current ratio can be obtained with the minimum supply voltage. However, semiconductor physics places a fundamental theoretical limit on the maximum slope at 60 mV/decade at room temperature [6]. Exploiting negative capacitance in NC-FET can presumably overcome this limit to achieve a class of low-power transistors with a steep subthreshold slope sharper than 60 mV/decade.

In the ferroelectric-based NC-FET, the starting assumption is the negative capacitance from the section with a negative slope in the *P-E* (polarization-field) curve shown in Fig. 1(a), which arises from the thermodynamic consideration of electrically induced transition from one polarization state to the opposite one. In this paper, we discuss the transient electrical behavior of the ferroelectric-based NC-FET.

A conventional *P-E* hysteresis curve typically measured in ferroelectric capacitors (Fig. 1(b)) does not exhibit a negative slope, and hence, in contrast to the loop in Fig. 1(a), has no negative capacitance. The schematics shown in Fig. 2 help illustrate this difference and the benefits of negative capacitance. The NC-FET structure comprises a ferroelectric layer added on top of the gate dielectric of a regular MOSFET. The gate voltage $V_g$ applied in the direction of turning the transistor ON drops partially across the ferroelectric, $V_{FE}$, and partially across the MOSFET, $V_{FET}$, so that $V_g = V_{FE} + V_{FET}$. With an incremental gate bias $\Delta V_g$, the applied field induces polarization within the ferroelectric. Note that the depolarization field, associated with the switched polarization, acts against the applied field. The net charge density, defined as a difference between the polarization $P$ and the screening charge density $\sigma$, determines the net field inside the ferroelectric capacitor, and ultimately the terminal voltage across it $\Delta V_{FE}$. Negative capacitance implies that the switched polarization is not completely screened during switching so that $|P| > |\sigma|$. From

$$\Delta V_{FE} = \frac{(\sigma - P)t_{FE}}{\varepsilon}, \qquad (1)$$

($\varepsilon$ is permittivity and $t_{FE}$ is ferroelectric thickness) $\Delta V_{FE}$ is negative and that $\Delta V_{FET} > \Delta V_g$ (from $\Delta V_g = \Delta V_{FET} + \Delta V_{FE}$). The last inequality is the voltage gain that NC-FET is in pursuit of, which leads to steeper subthreshold slope and higher current.

We now focus on the transient response of both the polarization and screening charges. As was pointed out by Catalan et al [5], the polarization process can occur at a much faster rate than the response of the screening charges. While the polarization switching time $t_s$ can be in the sub-ns range (which is still well above the ultimate physical limit) [7,8], the screening charges need much longer time to re-arrange for a new polarization state. Specifically, redistribution of the screening charges on the electrodes occurs with the characteristic time $\tau_{sc}$ determined by many factors; the dielectric properties of the ferroelectrics and interfaces, the electrode materials, and, especially, the external circuitry parameters. This time is typically of the order of several nanoseconds or longer [9]. Hence, negative capacitance can be detected only when the polarization has already switched (or at least partially switched) but the screening has not completed yet. To help measure negative capacitance, it was suggested to slow down the screening process by adding a large resistor in series to increase $\tau_{sc}$ [1].

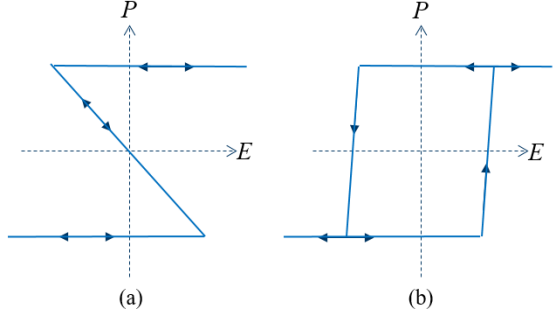

Fig. 1. *P-E* curve across the ferroelectric terminals; (a) with negative capacitance and (b) without negative capacitance and with hysteresis.

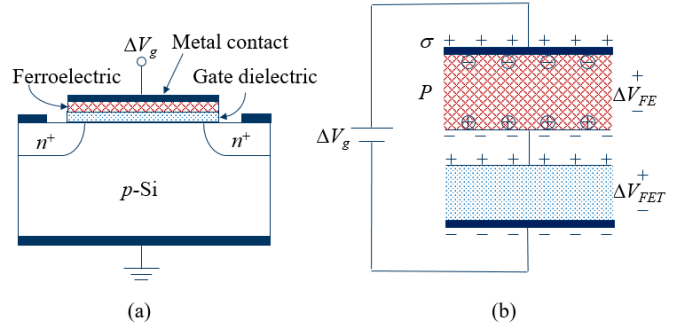

Fig. 2. (a) Schematics of NC-FET with a ferroelectric on top of a regular MOSFET. (b) Voltage and charge distributions in NC-FET. Circled charges *P* are due to polarization. Un-circled charges $\sigma$ are screening charges on the electrodes or at the surfaces/interfaces. Internal screening charges inside the ferroelectric are ignored. $\Delta V_{FE}$ is voltage across the ferroelectric. $\Delta V_{FET}$ is voltage across the rest of the structure, which is similar to a regular MOSFET, and $\Delta V_g = \Delta V_{FE} + \Delta V_{FET}$.

Figure 3 illustrates the transient behavior of the NC-FET qualitatively. After a step voltage is applied to the gate to turn on the *n*-channel transistor (Fig 3(a)), it takes a certain switching time $t_s$ for polarization to reverse (Fig. 3(b)), and redistribution of the screening charges follows with a larger time constant $\tau_{sc}$. Figure 3(c) shows the change of voltage across the ferroelectric capacitor $\Delta V_{FE}$ during these processes. The negative voltage value comes from the polarization *P* before the new distribution of the screening charges $\sigma$ is fully established. As the screening charges begin to balance out the polarization charges, the negative voltage across the ferroelectric diminishes. Figure 3(d) shows that the NC-FET current is higher than that of a regular MOSFET, but only up until $\tau_{sc}$. Thus, an important point to be emphasized here is that in steady state the ON-current cannot capitalize on the negative capacitance effect. Equivalently, the *P-E* loop in Fig. 1(a) is only transient and transforms into the loop in Fig. 1(b) with time.

As emphasized in Fig. 4, the steady-state subthreshold slope and the ON-current are likely to be lower than those of a MOSFET, due to the added ferroelectric layer leading to a thicker total dielectric thickness. But a faster gate-voltage ramp can help to realize larger current and steeper subthreshold slope. This can explain some of the positive measurements reported.

It has been proposed to add positive capacitance in series to "stabilize" the negative capacitance. It is necessary to balance the composite system to have positive net capacitance to be "stabilized". Also, this series capacitor is acting as a load line with the proper value and slope to intercept the *P-E* curve in Fig. 1(a) at the region of negative slope [2]. However, while it is indeed a necessary condition for NC-FET to operate, it addresses different concerns and does not stabilize the transient effects (in time evolution) discussed here.

It would be beneficial here to compare the different time scales of NC measurements and NC-FET device operation. Commonly in the *I-V* measurements, including that of the subthreshold $\log(I_d)$-$V_g$ curves, the voltage ramping rates are in the order of 10 - 100 ms. In the capacitance measurements, a small AC signal time domain usually lies in the range from 1 μs to 0.1 ms (frequency domain from 1 MHz to 10 kHz). These time constants are much longer than the typical $\tau_{sc}$. Hence, our point is that negative capacitance and transistor subthreshold slope steeper than the 60 mV/decade cannot be realized in DC (steady-state) conditions. Faster measurements are a better approach to observe this phenomenon. It is also important to fully characterize $\tau_{sc}$, in addition to $t_s$, as they both have important effect on circuit performance as briefly discussed later.

Some examples of measurements beneficial to get some idea of the screening time include subthreshold characteristics measured with different gate voltage ramping rates. Another is a step voltage to the gate from $V_{g1}$ to $V_{g2}$, and monitor the response of FET current as a function of time.

While we empathize that faster gate voltage ramp can help realize negative capacitance and hence steeper subthreshold slope, one has to also monitor the hysteresis between the ramping up (turning on) and ramping down (turning off) directions. The hysteresis can be worse with faster ramping rate, and the over-all effects have to be addressed.

For optimum circuit performance, the polarization switching time $t_s$ should be as short as possible, as it sets the high speed limit of using the negative capacitance feature. On the other hand, the screening time $\tau_{sc}$ should be as long as possible for the benefit of higher current. Specifically, in a CMOS inverter, $\tau_{sc}$ should be long enough to discharge (for *n*-channel device) all the nodal charges to reach the low-voltage state. This requires accurate knowledge of $\tau_{sc}$, the circuit topology, and the parasitic effects. In this respect, accurate time-dependent compact models would be necessary to get the detailed performance of different circuits. But the NC-FET compact models [10 -11] need to capture the screening-charge-related transient characteristics and even the hysteresis effects, so accurate characterization of the transient response of NC-FET is mandatory.

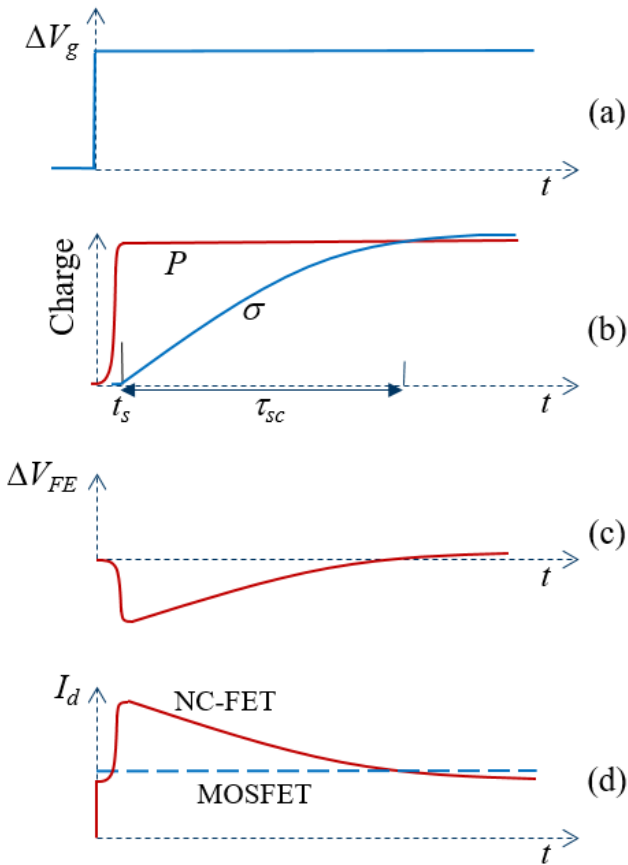

Fig. 3. Qualitative transient characteristics of NC-FET: (a) applied gate bias for turn-ON; (b) polarization charges $P$ within the ferroelectric and screening charges $\sigma$ in electrodes; (c) net voltage across the ferroelectric capacitor; and (d) drain current of NC-FET in comparison to a regular MOSFET. Note that a regular MOSFET has a faster response (without $t_s$) and larger current in the steady state.

There are additional issues to be dealt with if the NC-FETs are to be used in real circuits. First, reliability needs to be checked since the gate dielectric in series with the ferroelectric will experience a higher field than produced by the supply voltage, due to the voltage gain. Second, in this paper, we have discussed the turn-ON process, but the turn-OFF cycle needs to be examined as well. As discussed, the transistor current at a steady state will likely be lower than that of a regular MOSFET, and this impact will depend on circuit topology. The requirement of a matching capacitance to that of the ferroelectric layer may also demand a thicker oxide layer than state-of-the-art in MOSFETs. Another well-known potential problem is hysteresis between turning-on and turning-off. For analog circuit applications, all these issues are expected to be more complicated and critical.

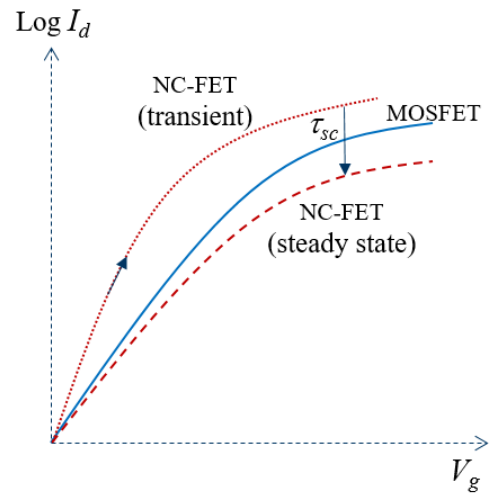

Fig. 4. Log($I_d$)-$V_g$ curves of FETs, comparing those of NC-FET in transient (dotted line), in the steady state (dash line), and of regular MOSFET (solid line). Curves or threshold voltages are normalized (shifted horizontally) for the same OFF-current.

In summary, we analyzed the transient nature of negative capacitance in ferroelectric devices, which stems from the difference in the rates of polarization reversal and screening processes. In order to observe and utilize the steep subthreshold in NC-FET devices, we suggest that one has to perform pulse measurements in a time scale much shorter in comparison with the polarization switching time. More importantly, application of NC-FETs needs to be examined in the context of real-circuit performance where the NC transient nature is taken into account. One of the attractive approaches for realization of the NC-FET devices could be employment of the multidomain ferroelectric superlattices [3] where polarization screening is inherently suppressed, potentially giving rise to a more steady negative capacitance effect.

Acknowledgments

A.G. acknowledges the support by the Center for Nanoferroic Devices (CNFD), a Semiconductor Research Corporation Nanoelectronics Research Initiative (SRC-NRI) under Task ID 2398.002, sponsored by NIST and the Nanoelectronics Research Corporation (NERC).

REFERENCES

[1] A. Khan, K. Chatterjee, B. Wang, S. Drapcho, L. You, C. Serrao, S. Bakaul, R. Ramesh, and S. Salahuddin, "Negative capacitance in a ferroelectric capacitor", *Nature Mater.*, **14**, 182 (2015).
[2] S. Salahuddin and S. Datta, "Use of Negative Capacitance to Provide Voltage Amplification for Low Power Nanoscale Devices", *Nano Lett.*, **8**, 405 (2008).
[3] P. Zubko, J. Wojdeł, M. Hadjimichael, S. Fernandez-Pena, A. Sené, I. Luk'yanchuk, J. Triscone, and J. Íñiguez, "Negative capacitance in multidomain ferroelectric superlattices", *Nature*, **534**, 524 (2016).
[4] C. Krowne, S. Kirchoefer, W. Chang, J. Pond, and L. Alldredge, "Examination of the possibility of negative capacitance using ferroelectric materials in solid state electronic devices", *Nano Lett.*, **11**, 988 (2011).
[5] G. Catalan, D. Jimenez, and A. Gruverman, "Negative capacitance detected", *Nature Mater.*, **14**, 137 (2015).

[6] S. Sze and K. Ng, *Physics of Semiconductor Devices*, 3rd Ed., Wiley, 2006.

[7] J. Li, B. Nagaraj, H. Liang, W. Cao, C. Lee, and R. Ramesh, "Ultrafast polarization switching in thin-film ferroelectrics", *Appl. Phys. Lett.*, **84**, 1174 (2004).

[8] J. Scott and C. Araujo," Ferroelectric memories", *Science*, **246**, 1400 (1989).

[9] P. Larsen, G. Kampschöer, M. Ulenaers, G. Spierings, and R. Cuppens, "Nanosecond switching of thin ferroelectric films", *Appl. Phys. Lett.*, **59**, 611 (1991).

[10] G. Pahwa, T. Dutta, A. Agarwal, S. Khandelwal, S. Salahuddin, C. Hu, and Y. Chauhan, "Analysis and Compact Modeling of Negative Capacitance Transistor with High ON-Current and Negative Output Differential Resistance—Part I: Model Description", *IEEE Trans. Electron Devices*, **63**, 4981 (2016).

[11] G. Pahwa, T. Dutta, A. Agarwal, S. Khandelwal, S. Salahuddin, C. Hu, and Y. Chauhan, "Analysis and Compact Modeling of Negative Capacitance Transistor with High ON-Current and Negative Output Differential Resistance—Part II: Model Validation", *IEEE Trans. Electron Devices*, **63**, 4986 (2016).